\documentstyle[12pt]{article}
\begin{document}
\title{Photon-Photon Scattering, Pion Polarizability and Chiral Symmetry}
\author{John F. Donoghue and Barry R. Holstein\\[5mm]
Department of Physics and Astronomy\\
University of Massachusetts\\
Amherst, MA 01003}
\begin{titlepage}
\maketitle
\begin{abstract}

Recent attempts to detect the pion polarizability via analysis of
$\gamma\gamma\rightarrow\pi\pi$ measurements are examined. The connection
between calculations based on dispersion relations and on chiral perturbation
theory is established by matching the low energy chiral amplitude with
that given by a full dispersive treatment.  Using the values for the
polarizability required by chiral symmetry, predicted and experimental
cross sections are shown to be in agreement.
\end{abstract}
{\vfill UMHEP-383}
\end{titlepage}

\section{Introduction}

\indent The reactions $\gamma\gamma\rightarrow\pi^0\pi^0$ and
$\gamma\gamma\rightarrow\pi^+\pi^-$ represent currently interesting theoretical
and experimental laboratories for chiral perturbation theory ($\chi$PT)\cite{1}
and for
dispersion relations\cite{2}. For charged pion production the
$\chi$PT prediction is in
good agreement with the data, as shown in Figure 1\cite{3,4}.
However, in the case of
neutral pion production, the one loop chiral perturbation theory calculation
\cite{4,5} disagrees even near threshold, with both a dispersive
treatment and the data, as can be seen in
Figure 2\cite{6}.  \footnote{The dispersive prediction shown therein is that
which we describe later in this paper, but it is similar to the pioneering
dispersive calculation performed by Morgan and Pennington\cite{7}.}
This situation at first appears surprising, as a dispersive
calculation should
obey the chiral strictures at low energy, while a chiral calculation should
obey
the unitarity properties to the order in energy that one is working.  Of
course,
the chiral result is known to be expansion in the energy, and it is always
possible for higher orders to modify the first order result\cite{8}.
However, in most
other calculations the modifications are not very large near threshold.  One of
the purposes of this paper is to resolve the theoretical issue of the
connection
between the chiral and dispersive methods, and to understand the origin of
large
corrections to the $\gamma\gamma\rightarrow\pi^0\pi^0$ amplitude near
threshold.  We do this in Section II by matching the two descriptions, and
providing an analytic solution to the dispersion relation which is consistent
with the low energy chiral properties.  This exercise indicates that the two
descriptions are in fact completely consistent in their respective limits, and
suggests that rescattering effects required by unitarity are the dominant
source
of corrections to the lowest order chiral prediction.

In addition, the two photon reactions have been utilized phenomenologically in
order to extract the pion electromagnetic polarizability\cite{9}.  For this
purpose,
one needs as accurate a description of the amplitude as possible, and we use
our
results from Section II to construct an improved picture of the transition
amplitude in Section III.  The
connection with the polarizability and a review of the present experimental
status is given in Section IV, and our results are summarized in a concluding
Section V.

\section{Matching the chiral and dispersive descriptions}

\indent In this section our main interest is to understand how a dispersive
treatment
matches on to the calculation of chiral perturbation theory and to learn why
there exist large corrections to the chiral results even near threshold.  In a
recent paper, Pennington has run a series of numerical exercises which suggest
that the necessary modifications come from multiloop effects, which are of
higher order in the chiral expansion\cite{10}. Our analytic study, discussed
below,
confirms this conclusion. In fact the results turn out to be quite simple, and
we will be able to neatly identify the source of the corrections.

We begin by setting up a bit of formalism.  We shall assume, consistent with
the
chiral expansion, that when we are in the near-threshold region the only
relevant higher order effects are in the helicity conserving S-channel
amplitude,
which we write as
\begin{eqnarray}
& &\gamma\gamma\rightarrow\pi^{+}\pi^{-}:\qquad
f^{C}(s)=\frac{1}{3}\left[2f_{0}(s) +
f_{2}(s)\right]\nonumber\\
& &\gamma\gamma\rightarrow\pi^{0}\pi^{0}:\qquad
f^{N}(s)=\frac{2}{3}\left[f_{0}(s) -
f_{2}(s)\right]
\end{eqnarray}
where I = 0, 2 refers to the isospin of the final $\pi\pi$ state.  For neutral
pion production and working in the gauge wherein $\epsilon_{2}\cdot k_{2} =
\epsilon_{2} \cdot k_{1} = \epsilon_{1} \cdot k_{2} = \epsilon_{1} \cdot k_{1}
=
0$ the transition amplitude is
\begin{equation}
\gamma\gamma\rightarrow\pi^{0}\pi^{0}:\qquad  {\rm Amp} = 2ie^{2} \epsilon_{1}
\cdot
\epsilon_{2} f^{N}(s)
\end{equation}
and the cross section is given by
\begin{equation}
\frac{d\sigma^{N}}{d\Omega} = \frac{\alpha^{2}}{4s} \beta(s)|f^{N}(s)|^{2}
\end{equation}
where
\begin{equation}
\beta(s) = \sqrt{\frac{s-4m_{\pi}^{2}}{s}}
\end{equation}
is the center of mass velocity of the produced pions.  For convenience in
comparison with experimental results it is useful to present also the total
cross
section for events having $|\cos\theta|$ less than some fixed value Z
\begin{equation}
\sigma(|\cos\theta|<Z) = \frac{\pi\alpha^{2}Z}{s} \beta(s) |f^{N}(s)|^{2}.
\end{equation}
In the charged pion case the Born and seagull contributions to this multipole
must also be included, so that the full amplitude becomes
\begin{eqnarray}
\gamma\gamma\rightarrow\pi^{+}\pi^{-}:\qquad {\rm Amp} =
2ie^{2}\left[\epsilon_{1}\cdot\epsilon_{2} a(s)
- \frac{\epsilon_{1}\cdot p_{+} \epsilon_{2}\cdot p_{-}}{p_{+} \cdot k_{1}} -
\frac{\epsilon_{1} \cdot p_{-}\epsilon_{2} \cdot p_{+}}{p_{+} \cdot
k_{2}}\right]
\end{eqnarray}
with cross section
\begin{equation}
\frac{d\sigma}{d\Omega} = \frac{\alpha^{2}}{2s} \beta(s)\left[|a(s)|^{2} -
2{\rm Re} a(s)
\frac{\beta^{2}(s)\sin^{2}\theta}{1-\beta^{2}(s)\cos^{2}\theta} + 2
\frac{\beta^{4}(s)\sin^{4}\theta}{\left(1-\beta^{2}(s)\cos^{2}\theta\right)^{2}}\right]
\end{equation}
Here
\begin{equation}
a(s) = 1 + f^{C}(s) - f^{C}_{\rm Born}(s)
\end{equation}
where
\begin{equation}
f^{C}_{\rm Born}(s) = \frac{1-\beta^{2}(s)}{2\beta(s)} ln
\left(\frac{1+\beta(s)}{1-\beta(s)}\right) = f^{\rm Born}_{0}(s) = f^{\rm
Born}_{2}(s)
\end{equation}
is the Born approximation value for the helicity conserving S-wave multipole.
Again in order to compare with data we integrate Eq. 7 to yield
\begin{eqnarray}
\sigma (|\cos \theta | )< Z)={\pi \alpha^2 \beta (s) \over s}\left[ 2Z\left(
|a|^2+2-2{\rm Re}a+{(1-\beta ^2(s))^2\over 1-\beta ^2(s)Z^2}\right)
\right. \nonumber\\
\left. +{1-\beta ^2(s) \over \beta (s)} ln \left( {1+\beta(s)Z\over 1-\beta
(s)Z}
\right) (2{\rm Re}a-3-\beta ^2(s))\right]
\end{eqnarray}
In the threshold region the phase of $f_{I}(s)$ is required by unitarity to
be equal to the corresponding $\pi\pi$ phase shift $\delta_{I}(s)$.  When
$s>16m^{2}_{\pi}$, inelastic reactions involving four pions are allowed.
However,
the inelasticity is small, being of order $E^{8}$ in the chiral expansion and
also suppressed by phase space considerations. In practice, the inelasticity is
negligible up to $K\bar{K}$ threshold, $s\sim 1GeV^{2}$,and consequently we
will
neglect inelasticity throughout our analysis.

The functions $f_{I}(s)$ are then analytic functions of s except for cuts along
the positive and negative real axis. For positive $s$, the right hand cut
extends
from $4m^{2}_{\pi} < s < \infty$ and is due to the $s$ channel $\pi\pi$ state.
For
negative $s$, the left hand cut is due to $t,u$-channel intermediate states
such
as $\gamma\pi\rightarrow\pi\rightarrow\gamma\pi$ or
$\gamma\pi\rightarrow\rho\rightarrow\gamma\pi$, and extends from $-\infty<s<0$.

The single channel final state unitarization problem has a simple solution in
terms of the Omnes function\cite{11}
\begin{equation}
D_I^{-1}(s)=\exp\left({s\over \pi}\int^\infty_{4m_\pi^2} {ds'\over s'}
{\delta_I(s')\over s'-s-i\epsilon}\right)
\end{equation}
---the result must have the form
\begin{equation}
f_I(s)=g_I(s)D_I^{-1}(s)
\end{equation}
where $g_{I}(s)$ is an analytic function with no cuts along the position real
axis.  Morgan and Pennington consider a function $p_{I}(s)$ which has the same
left hand singularity structure as $f_{I}(s)$, but which is real for $s>0$.
They
then write a twice subtracted dispersion relation for the difference
$(f_{I}(s)-p_{I}(s))D_{I}(s)$, with the result\cite{7}
\begin{equation}
f_I(s)=D_I^{-1}(s)\left[ p_I(s)D_I(s)+(c_I+sd_I)-{s^2\over \pi}\int^\infty
_{4m_\pi^2}{ds'\over {s'}^2}{p_I(s'){\rm Im}D_I(s')\over s'-s-i\epsilon}\right]
\end{equation}
where $c_{I},d_{I}$ are subtraction constants.  The combination
inside the square brackets is real
with any $p_{I}(s)$ which is real for $s>0$. By Low's
theorem $f_{I}$ must reduce to the Born term at low energies\cite{12}
\begin{equation}
f_I(s)=f_I^{\rm Born}(s)+{\cal O}(s)=p_I(s)+c_I+d_Is+\cdots
\end{equation}
so that we can set $c_{I}=0$ if we choose $p_{I}(s)=f^{Born}_{I}+\sl{O}(s)$.
The
only assumption made thus far has been the neglect of inelastic channels.

Analyticity and unitarity do {\em not} determine the remaining subtraction
constants $d_{0},d_{2}$. However, by matching the dispersion relation with the
low
energy chiral representation one can express $d_{0},d_{2}$ in terms of known
chiral low energy constants. This methodology was developed in Ref. 13, and we
apply it here. At low energies we set
\begin{equation}
p_I(s)=f_I^{\rm Born}(s),\qquad {\rm Im}D_I(s)=-\beta (s)t_I^{\rm CA}(s)
\end{equation}
where $t^{\rm CA}_{I}(s)$ are the lowest order (Weinberg) $\pi\pi$ scattering
amplitudes\cite{14}
\begin{equation}
t_0^{\rm CA}(s)={2s-m_\pi^2 \over 32\pi F_\pi^2}, \qquad
t_2^{\rm CA}(s)=-{s-2m_\pi^2 \over 32\pi F_\pi^2}
\end{equation}
Since these are simple polynominals of the form $t^{\rm CA}_{I}(s)=a+bs$, the
dipersive integral can be done exactly\footnote{Note that both left- and
right-hand sides of this equation have
identical imaginary parts and behave as ${\cal O}(s^2)$ for $s\sim 0$.}
\begin{eqnarray}
&-&{s^2\over \pi}\int^\infty_{4m_\pi^2}{ds'\over {s'}^2}{1-\beta^2(s')\over
2\beta (s')}ln\left({1+\beta (s')\over 1-\beta (s')}\right)
{\beta (s')t_I^{\rm CA}(s')\over s'-s-i\epsilon}\nonumber\\
&=&{1-\beta^2(s)\over 4\pi}ln^2\left({\beta (s)+1\over \beta (s)-1}\right)
t_I^{\rm CA}(s)+{1\over \pi}t_I^{\rm CA}(s)+{s\over 12\pi m_\pi^2}
t_I^{\rm CA}(0)
\end{eqnarray}
which yields a representation for the scattering amplitude\cite{15}
\begin{eqnarray}
f_I(s)&\equiv &f_I^{\rm Born}(s)+g_I(s)\nonumber\\
&=&D_I^{-1}(s)\left[ D_I(s){1-\beta^2(s)\over 2\beta (s)}ln\left({1+\beta(s)
\over 1-\beta (s)}\right) \right.\nonumber\\
& &-{1\over 4\pi}(1-\beta^2(s))t_I^{\rm CA}(s)ln^2
\left( {\beta (s)+1\over \beta (s)-1}\right)\nonumber\\
& &\left. -{1\over \pi}t_I^{\rm CA}(s)+s(d_I-{t_I^{\rm CA}(0)\over 12\pi
m_\pi^2})
+\Delta_I(s)\right].
\end{eqnarray}
Here $\Delta_{I}(s)$ represents the remainder which accounts for the difference
in the true dispersion integral from the lowest order inputs given in Eq. 15.
At
low energies $\Delta_{I}(s)\sim{\cal O}(s^{2})$.  Eq. 18 can be compared with
the
one loop $\sl{O}(E^{4})$ chiral amplitude which has the form\cite{4}
\begin{eqnarray}
f_I^{\rm Chiral}(s)&=&{1-\beta^2(s)\over 2\beta (s)}ln\left({1+\beta (s)
\over 1-\beta (s)}\right)-{1-\beta^2(s)\over 4\pi}t_I^{\rm CA}(s)
ln^2\left( {\beta (s)+1\over \beta (s)-1}\right)\nonumber\\
&-&{1\over \pi}t_I^{\rm CA}(s)+{2\over F_\pi^2}(L_9^r+L_{10}^r)s+\cdots
\end{eqnarray}
where $L^r_{9}+L^r_{10}$ is a combination of known chiral low energy
constants. This combination is independent of the renormalization scale and has
magnitude\cite{16}
\begin{equation}
L_9^r+L_{10}^r=(1.43\pm 0.27)\times 10^{-3}
\end{equation}
determined from radiative pion decay. Thus chiral symmetry fixes unambiguously
the subtraction constants which appear in the dispersive analysis---
\begin{equation}
d_I={2\over F_\pi^2}(L_9^r+L_{10}^r)+{t_I^{\rm CA}(0)\over 12\pi m_\pi^2}
\end{equation}
The two formalisms match very nicely at low energy yielding a parameter-free
descriptive of the low energy $\gamma\gamma\rightarrow\pi\pi$ process.

At this stage we can inquire into the origin of the large corrections found in
the $\gamma\gamma\rightarrow\pi^{0}\pi^{0}$ amplitude. Do they arise simply
from
the unitarization of the amplitude ({\it i.e.} $D_{I}(s)\neq 1$) or are new
inputs
needed in the amplitude (in which case $\Delta_{I}(s)$ would be most
important)?
We will argue that the rescattering physics in $D^{-1}_{I}(s)$ is most
important,
and that the main corrections are due to well-known ingredients.  In the next
section, we will attempt a full phenomenological treatment but here let us
explore the case with $\Delta_{I}(s)=0$ and a simple analytic form for
$D^{-1}_{I}(s)$.  The condition $Im D_{I}(s)=-\beta t^{\rm CA}_{I}(s)$ defines
the
[0,1] Pad\'{e} approximation for the Omnes function\cite{16}, {\it i.e.}
\begin{eqnarray}
D_I^{-1}(s)&=&{1\over 1-k_Is+t_I^{\rm CA}(s)(h(s)-h(0))}\nonumber\\
{\rm with} \qquad h(s)&=&{\beta (s)\over \pi}ln\left( {\beta (s) +1\over
\beta(s)
-1}\right), \qquad h(0)={2\over \pi}
\end{eqnarray}
and allows one an approximate but  simple analytic representation for the
$\gamma\gamma\rightarrow\pi\pi$ amplitude, so we will use this form in the
remainer of this section.  The constant $k_{0} \cong \frac{1}{25m_{\pi}^{2}}$
is
chosen to match the small $s$ behavior of the experimental $D^{-1}_{0}(s)$
function, and $k_{2}\cong -\frac{1}{30m_{\pi}^{2}}$ is chosen from a fit to $I
=
2$ $\pi\pi$ scattering.  For more details of both of these ingredients, see
Section
III.  The resulting form for the $\gamma\gamma\rightarrow\pi^{0}\pi^{0}$
amplitude is
\begin{eqnarray}
f^N(s)&=&-{1\over 48\pi^2F_\pi^2}\left( 1+{m_\pi^2\over s}ln^2\left(
{\beta (s)+1\over \beta (s)-1}\right)\right)\nonumber\\
&\times &\left[(2s-m_\pi^2)D_0^{-1}(s)+
(s-2m_\pi^2)D_2^{-1}(s)\right]\nonumber\\
&+&{4\over 3F_\pi^2}(L_9^r+L_{10}^r)s
(D_0^{-1}(s)-D_2^{-1}(s))
\end{eqnarray}
which, when the Pad\'{e} forms of $D^{-1}_I(s)$ are used, provides a
consistent analytic solution to the dispersion relation while also displaying
the
correct chiral properties to ${\cal O}(s)$.  In Figure 3, we plot the resulting
cross section, in
comparison with the data and the lowest order result. It can be seen that the
Omnes functions produce a substantial modification even near threshold. Of
these,
the most important is $D^{-1}_{0}(s)$ which reflects the strong attractive
$\pi\pi$ scattering in the $I=0, J=0$ channel\cite{18}.  The use of an
empirical determination of
$D^{-1}_{0}(s)$ in the next section will further increase the amplitude. While
refinements can be added to the calculation of the amplitude, we conclude that
the major ingredient which modifies the threshold behavior in
$\gamma\gamma\rightarrow\pi^{0}\pi^{0}$ is the final state rescattering
corrections.

That such corrections might be important is perhaps in retrospect not so
surprising.  Chiral perturbation theory represents an expansion in energy with
a
scale of order $\Lambda_{\chi}\sim 4\pi F_{\pi}\sim1GeV$\cite{19}.
For center of mass
energies $\sqrt{s} \leq 0.5 GeV$ one would expect that $\chi$PT should give an
accurate representation of the scattering amplitude, and this is indeed the
case
for the $\gamma\gamma\rightarrow\pi^{+}\pi^{-}$ process.  However, for
$\gamma\gamma\rightarrow\pi^{0}\pi^{0}$ there exist no Born or $\sl{O}(E^{4})$
counterterm contributions.  The $\sl{O}(E^{4})$ amplitude arises entirely from
one
loop effects and is consequently nearly an order of magnitude smaller than its
charged pion counterpart. It is this smallness which accounts for the
importance
of higher order effects, and one should be alerted to the possible significance
of such corrections in other such processes such as $K_{S}\rightarrow2\gamma,
K_{L}\rightarrow\pi^{0}\gamma\gamma, \eta^{0}\rightarrow\pi^{0}\gamma\gamma$
etc.

\section{Further refinement}
\indent The analysis of the previous section was done in a particularly naive
limit in
order to expose the essential physics in the clearest fashion. Although this
provides a good description of the threshold region, in phenomenological
studies
one may be interested in a more complete calculation. We provide this in the
present section. In particular we add the following ingredients:

i) The Omnes function $D^{-1}_{0}(s)$ has been determined from the experimental
phase
shifts by Gasser et al.\cite{13}. We use this in place of the Pad\'{e}
approximation
Eq. 22.

ii) The Born amplitude is not sufficient to fully describe the
$\gamma\gamma\pi\pi$ vertex which receives further contributions from
$\rho,\omega,A1$ exchanges. We add these to the formalism. The resulting
amplitude
is similar to that of Morgan and Pennington\cite{9} with the exception of the
related
ingredients of $L_{9}^r+L_{10}^r$ and A1 exchange.  As we describe more fully
below,
the A1 contribution is in fact more important at low energy than is the effect
of
the $\rho$ and $\omega$.

The Omnes function involves an integral over the $\pi\pi$ scattering phase
shifts. These are known experimentally up to above $1GeV$, and at low $s$ the
Omnes function is not very sensitive to the phase shifts beyond their known
range. Gasser et al. have taken this data, added chiral constraints at low
energy where the data is somewhat poor and performed a numerical evaluation
of $D^{-1}_{0}(s)$\cite{20}.  The result is somewhat larger in both the real
and imaginary
parts than the Pad\'{e} approximation used in the last section. The form of the
$I=2$ Omnes function is not as important because $D^{-1}_{2}(s)$ remains closer
to
unity. In this case we use the Pad\'{e} form, given in Eq. 22.  The constant
$k_{2}$ is  chosen so that the $\pi\pi$ scattering amplitude, defined by
\begin{equation}
t_2(s)=t_2^{\rm CA}(s)D_2^{-1}(s)
\end{equation}
matches the experimental phase shifts over the region $4m^{2}_{\pi}\leq s \leq
1GeV^{2}$.  We find that the constant $k_{2}=-{1\over 30m_\pi^2}$ provides a
good fit throughout
this region. We note that Gasser and Meissner compare the $I=0$ Pad\'{e}
approximation with the full Omnes function and with a two loop chiral
calculation
and find that the Pad\'{e} form does not completely characterize the chiral
logs properly\cite{21}. While it is important to keep in mind that the Pad\'{e}
procedure is only an approximation, the numerical differences are not large if
the free parameter is chosen properly. We expect that in the $I=2$ channel the
Pad\'{e} form should be numerically a good approximation.  Use of these
result in the dispersive formula of Eq. 13, with $p_{I}(s) = f^{Born}_{I}(s)$,
yields the curve in Figure 4.

The $\gamma\pi\rightarrow\gamma\pi$ Compton amplitude receives important
modification at low energy from $\rho$ and $A1$ exchange, as shown in Figure 5.
These have been analysed in detail in Ref. 22.  In particular, the Compton
amplitude
including these poles in a vector dominance model is given by
\begin{eqnarray}
& &{1\over 4\pi\alpha}T_{\mu\nu}(p_1,q_1,q_2)=2g_{\mu\nu}\nonumber\\
& &-{T_\mu(p_1,p_1+q_1)T_\nu(p_2+q_2,p_2)\over
(p_1+q_1)^2-m_\pi^2}-{T_\nu(p_1,p_1-q_2)T_\mu(p_2-q_1,p_2)\over (p_1-q_2)^2
-m_\pi^2}\nonumber\\
& &+{F_V^2\over F_\pi^2}\left( g_{\mu\nu}\left( {q_2^2\over
m_V^2-q_2^2}+{q_1^2\over m_V^2-q_1^2}\right) -q_{1\mu}q_{1\nu}{1\over m_V^2
-q_1^2}-q_{2\mu}q_{2\nu}{1\over m_V^2-q_2^2}\right)\nonumber\\
& &-{F_A^2\over F_\pi^2}(g_{\mu\nu}q_1\cdot q_2-q_{1\nu}q_{2\mu})
\left[ {1-{p_1\cdot(p_1+q_1)\over m_A^2}\over m_A^2-(p_1+q_1)^2}+
{1-{p_1\cdot(p_1-q_2)\over m_A^2}\over m_A^2-(p_1-q_2)^2}\right]
\nonumber\\
& &{\rm with}\nonumber\\
& & T_\mu(p_1,p_2)=(p_1+p_2)_\mu\nonumber\\
& &+{F_V^2\over 2F_\pi^2(m_V^2-q^2)}
\left[ (p_1+p_2)_\mu q^2-(p_1-p_2)_\mu (p_1^2-p_2^2)\right]
\end{eqnarray}
This is to be compared with the chiral form of the amplitude
\begin{eqnarray}
& &{1\over 4\pi\alpha}T_{\mu\nu}(p_1,q_1,q_2)=2g_{\mu\nu}\nonumber\\
& &-{T_\mu (p_1,P_1+q_1)T_\nu (p_2+q_2,p_2)\over
(p_1+q_1)^2-m_\pi^2}-{T_\nu (p_1,p_1-q_2)T_\mu (p_2-q_1,p_2)\over
(p_1-q_2)^2-m_\pi^2}\nonumber\\
& &+{4\over F_\pi^2}L_9^r((q_1^2+q_2^2)g_{\mu\nu}-q_{1\mu}q_{1\nu}
-q_{2\mu}q_{2\nu})-{8\over F_\pi^2}(L_9^r+L_{10}^r)(q_1\cdot q_2 g_{\mu\nu}
-q_{2\nu}q_{1\nu})\nonumber\\
& &{\rm with}\quad T_\mu (p_1,p_2)=(p_1+p_2)_\mu(1+{2L_9^r\over F_\pi^2}q^2)
-(p_1-p_2)_\mu{2L_9^r\over F_\pi^2}(p_1^2-p_2^2)
\end{eqnarray}
We see from this that this parameter $L_{9}^r$ is due to $\rho$ exchange,
$L_{10}$
involves $\rho + A1$ while the combination $L_{9}^r+L_{10}^r$ is purely$A1$
exchange if
the KSFR relation, $m_{A}=\sqrt{2} m_{\rho}$, is used\cite{23}.  These features
are by now
well known.  Thus the lowest order form of these has already been included in
the
previous analysis through the constant $L_{9}^r+L_{10}^r$.  However, as the
energy is
increased the momentum dependence in the propagator becomes more important and
should be explicitely included.  There are also the effect of $\rho$ and
$\omega$
exchanges from Figure 6.  The $\rho\pi\gamma$ and $\omega\pi\gamma$ couplings
are
of the form
\begin{equation}
A(V\rightarrow\pi\gamma)={1\over 2}\sqrt{R_V} \epsilon^{\mu\nu\alpha\beta}
V_{\mu\nu}\epsilon_{\alpha} p_{\pi\beta}
\end{equation}
Because of the powers of momentum in the vertex, the effect of these diagrams
are
suppressed at low energy, being of order $E^{6}$ in the chiral expansion. It is
interesting that $A1$ exchange is more important than $\rho$ and $\omega$ at
low
energies, and this point has been overlooked in Ref. 9.  Thus the S-wave
projections to be used in the dispersive integral should be\cite{24}
\begin{eqnarray}
p_I(s)&=&f_I^{\rm Born}(s)+p_{AI}(s)+p_{\rho I}(s)+p_{\omega I}(s)\nonumber\\
p_{A0}&=&p_{A2}={(L_9^r+L_{10}^r)\over F_\pi^2}\left( {m_A^2-m_\pi^2\over \beta
(s)}
ln\left( {1+\beta (s)+{s_A\over s}\over 1-\beta (s) +{s_A\over s}}\right) +
s\right)\nonumber\\
p_{\rho 0}&=&-{3R_\rho \over 2}\left( {m_\rho^2\over \beta (s)}ln\left(
{1+\beta (s) +{s_\rho\over s}\over 1-\beta (s)+{s_\rho\over s}}\right)
-s\right)
\nonumber\\
p_{\rho 2}&=&0\nonumber\\
p_{\omega 0}&=&-{1\over 2}p_{\omega 2}=-{R_\omega\over 2}\left({m_\omega^2\over
\beta (s)}ln\left({1+\beta (s)+{s_\omega\over s}\over 1-\beta (s)+
{s_\omega\over s}}\right)-s\right)
\end{eqnarray}
where $s_{i}=2(m^{2}_{i}-m^{2}_{\pi})$ and $R_{\omega}=1.35 GeV^{-2},
{}R_\rho =0.12 GeV^{-2}$
are determined from the condition
\begin{equation}
R_V={6m_V^3\over \alpha}{\Gamma(V\rightarrow \pi\gamma)\over
(m_V^2-m_\pi^2)^3}.
\end{equation}
The connection with our previous
formalism may be found by taking the limit $s\rightarrow 0$, whereby we find
\begin{eqnarray}
\lim_{s\rightarrow 0}p_{A0}(s)&=&\lim_{s\rightarrow 0}p_{A2}(s)=
{2(L_9^r+L_{10}^r)\over F_\pi^2}s\nonumber\\
\lim_{s\rightarrow 0}p_{\rho 0}(s)&=&{\cal O}(s^2)\nonumber\\
\lim_{s\rightarrow 0}p_{\omega 0}(s)&=&-{1\over 2}\lim_{s\rightarrow 0}
p_{\omega 2}(s)={\cal O}(s^2)
\end{eqnarray}
We observe that the contributions from $\omega,\rho$ exchange is ${\cal
O}(s^{2})$
and is outside the original chiral expansion, as claimed, while that
from $A_{1}$
exchange accounts for the ${\cal O}(s)$ chiral contribution from
$L_{9}+L_{10}$.
Thus since this piece is automatically included in the $A_{1}$ exchange term,
we
must modify the associated subtraction constants to become
\begin{equation}
d_I={t_I^{\rm CA}(0)\over 12\pi m_\pi^2}
\end{equation}
The contribution of the vector meson exchange terms can then
be included by defining the general
Compton scattering amplitude as\cite{24}
\begin{eqnarray}
& &{1\over 16\pi\alpha}T_{\mu\nu}(p_1,q_1,q_2)\equiv
A(q_{2\mu}q_{1\nu}-q_1\cdot
q_2g_{\mu\nu})\nonumber\\
&+&B\left({p_1\cdot q_1p_1\cdot q_2\over q_1\cdot q_2}g_{\mu\nu}+p_{1\mu}
p_{1\nu}-{p_1\cdot q_1\over q_1\cdot q_2}q_{2\mu}p_{1\nu}
-{p_1\cdot q_2\over q_1\cdot q_2}q_{1\nu}p_{1\mu}\right)
\end {eqnarray}
The neutral pion production cross section then can be written as
\begin{equation}
\gamma\gamma\rightarrow \pi^0\pi^0: \quad \sigma(|\cos \theta| <Z)=
{\pi\alpha^2\over s^2}\int_{t_a}^{t_b}dt\left(|A^0s-m_\pi^2B^0|^2
+{|B^0|^2\over s^2}(m_\pi^4 -tu)\right)
\end{equation}
where
\begin{equation}
t_{b\atop a}=m_\pi^2-{1\over 2}s\pm {sZ\over 2}\beta (s)
\end{equation}
and
\begin{eqnarray}
& &sA^0=-{2\over 3}(f_0(s)-f_2(s))+{2\over 3}(p_0(s)-p_2(s))
-{s\over 2}\sum_{V=\rho ,\omega}R_V\left( {m_\pi^2+t\over t-m_V^2}+
{m_\pi^2+u\over u-m_V^2}\right) \nonumber\\
& &B^0=-{s\over 2}\sum_{V=\rho ,\omega}R_V\left( {1\over t-m_V^2}
+{1\over u-m_V^2}\right)
\end{eqnarray}
while for charged pion production
\begin{equation}
\gamma\gamma\rightarrow\pi^+\pi^-:\quad \sigma(|\cos \theta |<Z)=
{2\pi\alpha^2\over s^2}\int^{t_b}_{t_a}dt\left(|A^+s-m_\pi^2B^+|^2
+{|B^+|^2\over s^2}(m_\pi^4 -tu)\right)
\end{equation}
with
\begin{eqnarray}
sA^+&=&-{1\over 3}(2f_0(s)+f_2(s))+{1\over 3}(2p_0(s)+p_2(s))\nonumber\\
& &-{s\over 2}R_\rho
\left({m_\pi^2+t\over t-m_\rho^2}+{m_\pi^2+u\over u-m_\rho^2}\right)\nonumber\\
& &+m_A^2s{L_9^r+L_{10}^r\over F_\pi^2}\left( {1-{m_\pi^2+t\over 2m_A^2}\over
t-m_A^2}
+{1-{m_\pi^2+u\over 2m_A^2}\over u-m_A^2}\right)\nonumber\\
B^+&=&-\left( {1\over t-m_\pi^2}+{1\over u-m_\pi^2}\right)
-{sR_\rho\over 2}\left( {1\over t-m_\rho^2}+{1\over u-m_\rho^2}\right)
\end{eqnarray}
Note that here the functions $f_{I}$ employ the modified subtraction constant
Eq.
28 and that the dispersion integrals utilize the full function $p_{I}(s)$
defined
in Eq. 26.  The integration over $t$ is performed analytically while the
dispersive integration is done numerically.  This then is our final form and
yields results for neutral and charged pion production as shown in Figures
2 and 7.
Note that in the low energy region {\it both} cross sections
are in good agreement with the
experimental data.  We should not expect consistency in the higher energy
sector---$\sqrt{s}\geq 700 MeV$---as important resonant effects associated with
the $f_{0}(975),f_{2}(1270)$ have not been included\cite{25}.

\section{Pion Polarizability}
The electromagnetic polarizability is a fundamental property of an elementary
particle which measures its deformation in the presence of an external
electric/magnetic field\cite{26}.  In the case of an atomic system this
property can be
probed by detection of the effects induced by the interaction of an
electromagnetic signal with a "box" filled with such atoms.  An example is
provided by the recent measurements of the proton polarizability as a byproduct
of low energy Compton scattering measurements on a hydrogen target.  In the
case
of the pion, of course, an appropriate target is not available.  Nevertheless,
it
is possible to probe the pion polarizability by measureing the Compton
scattering
amplitude, just as in the nucleonic analog, by exploiting either the process of
radiative pion-nucleon scattering $\pi N\rightarrow\pi N\gamma$ (Figure 8a),
pion
photoproduction in photon-nucleon scattering $\gamma N\rightarrow\gamma N\pi$
(Figure 8b) or direct $\gamma\gamma\rightarrow\pi\pi$ measurements (Figure 8c).
None of these experiments is straightforward.  However, in the case of the
charged pion each has been used to measure the polarizability, yielding
somewhat
discrepant results
\begin{eqnarray}
a)\quad \bar{\alpha}_E^{\pi^+}&=&(6.8 \pm 1.4 \pm 1.2) \times 10^{-4}{ }{\rm
fm}^3
\cite{27}\nonumber\\
b)\quad \bar{\alpha}_E^{\pi^+}&=&(20 \pm 12)
\times 10^{-4}{ }{\rm fm}^3\cite{28}\nonumber\\
c)\quad \bar{\alpha}_E^{\pi^+}&=&(2.2 \pm 1.6)\times 10^{-4}{ }{\rm
fm}^3\cite{29}
\end{eqnarray}
For neutral pions, only the $\gamma\gamma\rightarrow\pi\pi$ reaction has been
employed, and separate analysis using very different assumptions have yielded
the
results
\begin{eqnarray}
d)\quad |\bar{\alpha}_E^{\pi^+}|&=& (0.69 \pm 0.07 \pm 0.04)\times 10^{-4}
{\rm fm}^3\cite{29}\nonumber\\
e)\quad |\bar{\alpha}_E^{\pi^+}|&=&(0.8 \pm 2.0) \times 10^{-4} {\rm fm}^3
\cite{30}
\end{eqnarray}
Such measurements are of particular interest in that, as we shall show, for
both
charged and neutral pions chiral symmetry makes what should be a very reliable
prediction for the size of the polarizability\cite{31}
\begin{eqnarray}
\bar{\alpha}_E^{\pi^+}&=&2.7 \times 10^{-4}{ }{\rm fm}^3\nonumber\\
\bar{\alpha}_E^{\pi^0}&=&-0.5\times 10^{-4}{ }{\rm fm}^3
\end{eqnarray}
and thus the possible discrepences indicated by the data are potentially very
significant.

In this section we examine the $\gamma\gamma\rightarrow\pi\pi$ process as a
probe
of the polarizability. In particular we have seen that dispersion relations
coupled with chiral perturbation theory can be used in order to obtain a very
accurate description of the $\gamma\gamma\rightarrow\pi\pi$ cross section in
the
region $\sqrt{s} < 1GeV$ and this analysis can be modified in order to provide
an
experimental measure of the polarizability.  The connection of the
Compton amplitude with the polarizability may be found by
noting that for a particle with electric/magnetic polarizability
$\alpha_{E}/ \beta_{M}$ the associated energy is
\begin{equation}
U=-{1\over 2}4\pi\bar{\alpha}_E {\bf E}^2-{1\over 2}4\pi \bar{\beta}_M {\bf
H}^2
\end{equation}
Since, using our choice of gauge ${\bf E} \sim -
i \omega \hat{\epsilon}$,
${\bf B} = i{\bf k} \times\hat{\epsilon}$ we can identify the polarizability in
terms of the low-energy (cross channel) Compton scattering amplitude via
\begin{equation}
{\rm Amp}=\hat{\epsilon}_1\cdot\hat{\epsilon}_2\left(-{\alpha\over m_\pi}
+4\pi\bar{\alpha}_E\right)+\hat{\epsilon}_1\times{\bf
k}_1\cdot\hat{\epsilon}_2\times
{\bf k}_24\pi\bar{\beta}_M+\cdots .
\end{equation}
We find then\cite{32}
\footnote{Note that these forms do not obey the conventional
stricture $\bar{\alpha}_E^\pi =-\bar{\beta}_M^\pi$ which obtains in the chiral
limit.\cite{33}  However, this condition {\it is} satisfied if we take
$m_\pi\rightarrow 0$
as can be seen from the relations\cite{32}
\begin{eqnarray}
\bar{\alpha}_E^{\pi^+}+\bar{\beta}_M^{\pi^+}&=&4\alpha m_\pi
{R_\rho\over m_\rho^2-m_\pi^2}\approx 0.064\{0.39\pm 0.04\} \times 10^{-4}{
}{\rm fm}^3
\nonumber\\
\bar{\alpha}_E^{\pi^0}+\bar{\beta}_M^{\pi^0}&=&4\alpha m_\pi
\sum_V {R_V\over m_V^2-m_\pi^2}\approx 0.76\{1.04\pm 0.07\}\times 10^{-4}{
}{\rm fm}^3.
\end{eqnarray}
Thus the violations of this condition arise from the vector meson exchange
contributions, which are ${\cal O}(E^6)$ in the chiral expansion\cite{1}.  It
is
also interesting to see that both of Eqns. 43 are positive in agreement with
the dispersion relation requirement
\begin{equation}
\bar{\alpha}_E+\bar{\beta}_M={1\over 2\pi^2}\int_0^\infty {d\omega
\sigma_{\rm tot}(\omega )\over \omega ^2},
\end{equation}
experimental evaluation of which gives the bracketed values indicated in
Eq. 43.\cite{9}}
\begin{eqnarray}
\bar{\alpha}_E^{\pi^+}&=&-\lim_{t\rightarrow m_\pi^2,s\rightarrow 0}
{2\alpha\over m_\pi}\left(A^+(s,t)+{t-3m_\pi^2\over s}\tilde{B}^+(s,t)\right)
=2.68\times 10^{-4}{ }{\rm fm}^3\nonumber\\
\bar{\beta}_M^{\pi^+}&=&\lim_{t\rightarrow m_\pi^2,s\rightarrow 0}
{2\alpha\over m_\pi}\left(A^+(s,t)+{t-m_\pi^2\over s}\tilde{B}^+(s,t)\right)
=-2.61\times 10^{-4}{ }{\rm fm}^3\nonumber\\
\bar{\alpha}_E^{\pi^0}&=&-\lim_{t\rightarrow m_\pi^2,s\rightarrow 0}
{2\alpha\over m_\pi}\left(A^0(s,t)+{t-3m_\pi^2\over s}B^0(s,t)\right)
=-0.50\times 10^{-4}{ }{\rm fm}^3\nonumber\\
\bar{\beta}_M^{\pi^0}&=&\lim_{t\rightarrow m_\pi^2,s\rightarrow 0}
{2\alpha\over m_\pi}\left(A^0(s,t)+{t-m_\pi^2\over s}B^0(s,t)\right)
=1.26\times 10^{-4}{ }{\rm fm}^3
\end{eqnarray}
where we have defined
\begin{equation}
\tilde{B}^+(s,t)\equiv B^+(s,t)-B^+_{\rm Born}(s,t).
\end{equation}
In terms of these definitions the $\gamma\gamma\rightarrow\pi\pi$ amplitudes
can
be parameterized as
\begin{eqnarray}
f_{\pi^+}(s)&=&{1-\beta^2(s)\over 2\beta (s)} ln\left( {1+\beta (s)\over
1-\beta (s)}
\right) +{m_\pi\over 4\alpha}(\bar{\alpha}_E^{\pi^+}-\bar{\beta}_M^{\pi^+})s
+{\cal O}(s^2)\nonumber\\
f_{\pi^0}(s)&=&{m_\pi\over
4\alpha}(\bar{\alpha}_E^{\pi^0}-\bar{\beta}_M^{\pi^0})s+
{\cal O}(s^2)
\end{eqnarray}
where the explicit form of the $\sl{O}(s^{2})$ terms can be read off from
Eq. 13.  (Note that in this case, unlike that of Compton scattering, the
photons are colinear in the center of mass so that electric and magnetic
polarizability terms cannot be separaated and always appear in the combination
$\bar{\alpha}_E-\bar{\beta}_M $.)  Then instead of using the chiral symmetry
requirements as input
we can modify the linear component of the Compton amplitude in order to gauge
the sensitivity of the $\gamma\gamma\rightarrow\pi\pi$ as a probe of pion
polarizability.  Results of such variation are shown in Figures 9 and 10
for charged and neutral production respectively.

In the former case, it is clear that the experimental cross section is in
good agreement with the chiral symmetry prediction $\bar{\alpha}_E=2.7\times
10^{-4}
{\rm fm}^3$.  However, even 100\% changes in this value are also consistent
with the low energy data, as is clear from Figure 9.  We conclude that
although $\gamma\gamma\rightarrow\pi^+\pi^-$ measurements certainly have the
potential to provide a precise value for the pion polarizability, the
statistical uncertainty of the present values does not allow a particularly
precise evaluation.  In this regard our conclusions are in agreement with
those derived from the one loop chiral amplitude, although the uncertainty
in $\bar{\alpha}_E^{\pi^+}$ quoted by Babusci et al. in Ref. 9
seems somewhat smaller than that indicated
in our analysis.  Both results, however, appear to be inconsistent with
the value $(6.8\pm 1.4 \pm 1.2)\times 10^{-4} {\rm fm}^3$ quoted in Ref. 27.

In the case of neutral pion production our predicted cross section is
also in good agreement with the low energy data and therefore also with the
chiral prediction for $\bar{\alpha}_E^{\pi^0}$.  However, as shown in Figure 10
there is very little sensitivity to the polarizability and even much
improved measurements will not change this situation.  In this regard
our conclusions are in strong disagreement with those of
Babusci et al.\cite{9} wherein
the one loop chiral analysis was used in order to produce a rather
precise value for the neutral polarizability---$|\bar{\alpha}_E^{\pi^0}|=
0.69\pm 0.07 \pm 0.04\times 10^{-4}{ }{\rm fm}^3$.  This is because, as shown
above, higher loop corrections to the one-loop chiral prediction as
given by the dispersive analysis make essential corrections to the lowest
order result and bring agreement with the low energy data without any
need to modify any of the input parameters.

\section{Conclusions}

\indent Above we have shown how analyticity can be combined with the strictures
of chiral symmetry in order to allow a no-free-parameter description of the
low energy ($E<0.5GeV$) $\gamma\gamma\rightarrow\pi\pi$ process.  Specifically,
building on the work of Morgan and Pennington,
a doubly subtracted dispersion relation for the helicity-conserving S-wave
amplitude, with the subtraction constants determined in terms of
known chiral counterterms, augmented by the Born values for other multipoles
has been shown to be in good agreement with experimental data for both the
charged---$\gamma\gamma\rightarrow\pi^+\pi^-$---and neutral---
$\gamma\gamma\rightarrow\pi^0\pi^0$---channels.  We have also shown how these
results can be used in order to experimentally determine values for the
pion polarizability $\bar{\alpha}_E^\pi$.  In the case of the charged pion
process,
wherein the overall shape of the cross section is dominated by the Born
contribution, the one loop chiral correction which determines the
polarizability is the leading correction term and additional (multi-loop)
effects required by analyticity are found to be small.  The value of the
charged pion polarizability $\bar{\alpha}_E^{\pi^+}$ determined thereby is
is good agreement with both the chiral symmetry prediction as well with
that determined in an earlier one-loop chiral analysis of the same data.
This value disagrees at the three sigma level with that found via
radiative pion scattering and calls strongly for a remeasurement of the
latter process, as currently proposed at Fermilab.  In the case of the
neutral pion reaction, there exists no Born term and the requirements
of analyticity as embodied in the dispersion analysis are found to make
a substantial modification to the the one-loop chiral prediction, which
involves considerable cancellation between $I=0,2$ production amplitudes---
the full dispersive calculation is considerably larger in the threshold
region than is its one-loop analog.  This feature is the origin of the recent
claim that the value of the neutral pion polarizability extracted from
a one-loop analysis of this data must be about 40\% larger than the chiral
requirement.  We have shown that this conclusion is incorrect---a full
dispersive calculation including the strictures of chiral symmetry is
in good agreement with the measured neutral pion cross-section.  Indeed,
we conclude from our work that there exists no evidence in either
$\gamma\gamma\rightarrow\pi^+\pi^-$ or $\gamma\gamma\rightarrow\pi^0\pi^0$
for violation of the chiral symmetry predictions for the pion polarizability.

\medskip

{\bf Acknowledgements}:  This work was supported in part by the National
Science Foundation.

\newpage
\begin{center}
{\bf Figure Captions}
\end{center}

Figure 1 - The data points shown are the ${\gamma\gamma\rightarrow
\pi^+\pi^-}$ cross section (with $|\cos\theta |<Z\equiv 0.6$) measured by the
MARK-II collaboration (Ref. 3).  The dashed curve is the Born
approximation prediction, while the solid line is that from one-loop
chiral perturbation theory. \\

\medskip

Figure 2 - The data points shown are the $\gamma\gamma\rightarrow
\pi^0\pi^0$ cross section (with $|\cos\theta |<Z\equiv 0.8$) measured by the
Crystal Ball collaboration (Ref. 4).  The dashed curve is the prediction
of one-loop chiral perturbation theory, while the solid curve is a
full no-free-parameter dispersive calculation, as described in the text.\\

\medskip

Figure 3 - Shown is the $\gamma\gamma\rightarrow\pi^0\pi^0$ cross
section predicted by one-loop chiral perturbation theory (dashed line)
and by the simple analytic Pad\'{e} solution to the dispersion
relations (solid line).\\

\medskip

Figure 4 - Shown is the $\gamma\gamma\rightarrow\pi^0\pi^0$ cross
section predicted by one-loop chiral perturbation theory (dashed line)
and by a dispersive treatment using an Omnes function generated from
experimental $\pi\pi$ phase shifts in the S-wave I=0 channel (solid
line).\\

\medskip

Figure 5 - Shown are $\rho $ and $A1 $ exchange diagrams which affect
the Compton scattering amplitude at ${\cal O}(p^4)$.\\

\medskip

Figure 6 - Shown are $\rho $ and $\omega $ exchange diagrams which affect
the Compton scattering amplitude at ${\cal O}(p^6)$.\\

\medskip

Figure 7 - Shown is the $\gamma\gamma\rightarrow\pi^+\pi^-$ cross section
predicted by the Born approximation (dashed line), by one-loop chiral
perturbation theory (dot-dashed line) and by a full dispersive treatment
as described in the text (solid line).\\

\medskip

Figure 8 - Indicated are the various ways of obtaining experimental
values for the pion polarizability---a) radiative pion-nucleon scattering;
b) pion photoproduction in photon-nucleon scattering; and c) direct
$\gamma\gamma\rightarrow\pi\pi$ measurements.\\

\medskip

Figure 9 - Shown is the $\gamma\gamma\rightarrow\pi^+\pi^-$ cross section
predicted by the full dispersive calculation and with $\bar{\alpha}_E^{\pi^+}
=2.8\times 10^{-4}{ }{\rm fm}^3$ (solid line), with $\bar{\alpha}_E^{\pi^+}=
4.2\times 10^{-4}{ }{\rm fm}^3$ (upper dotted line), and with $\bar{alpha}_E
^{\pi^+}=1.4\times 10^{-4}{ }{\rm fm}^3$ (lower dotted line).\\

\medskip

Figure 10 - Shown is the $\gamma\gamma\rightarrow\pi^0\pi^0$ cross section
predicted by the full dispersive calculation and with $\bar{\alpha}_E^{\pi^0}=
-0.5\times 10^{-4}{ }{\rm fm}^3$ (solid line), with $\bar{\alpha}_E^{\pi^0}=
-1.3\times 10^{-4}{ }{\rm fm}^3$ (upper dotted line), and with
$\bar{\alpha}_E^{\pi^0}=
+0.3\times 10^{-4}{ }{\rm fm}^3$ (lower dotted line).

\end{document}